# Structural evidence of disorder induced two-step melting of vortex matter in Co-intercalated NbSe$_2$ single crystals from superheating and supercooling

Somesh Chandra Ganguli, Harkirat Singh, Indranil Roy, Vivas Bagwe, Dibyendu Bala, Arumugam Thamizhavel and Pratap Raychaudhuri[1]

*Tata Institute of Fundamental Research, Homi Bhabha Road, Colaba, Mumbai 400005, India.*

Disorder induced melting, where the increase in positional entropy created by random pinning sites drives the order-disorder transition in a periodic solid, provides an alternate route to the more conventional thermal melting. Here, using real space imaging of the vortex lattice through scanning tunneling spectroscopy, we show that in the presence of weak pinning, the vortex lattice in a type II superconductor disorders through two distinct topological transitions. Across each transition, we separately identify metastable states formed through superheating of the low temperature state or supercooling of the high temperature state. Comparing crystals with different levels of pinning we conclude that the two-step melting is fundamentally associated with the presence of random pinning which generates topological defects in the ordered vortex lattice.

---

[1] pratap@tifr.res.in

## I. Introduction

The vortex lattice (VL) in a Type II superconductor, where interactions between vortices which favor an ordered state competes with the random pinning potential caused by crystallographic defects, has been widely studied as a model system to understand the order to disorder transition (ODT) in the presence of random pinning[1]. For a clean system it has been speculated that the ordered VL realized in a 3-dimensional superconductor could melt into a liquid state at a characteristic temperature/magnetic field. Unambiguous experimental evidence of such melting has been observed in layered High-$T_c$ cuprates[2,3,4,5], which is attributed to their large operating temperatures, short coherence length and high degree of anisotropy which renders the system susceptible to thermal fluctuations. In conventional superconductors, the existence of VL melting has remained controversial. In the presence of weak or moderate pinning several studies find signatures of a thermodynamic first-order ODT[6,7,8,9,10]. However, experimental investigations on extremely pure Nb single crystals did not find evidence[11,12] of VL melting below the upper critical field ($H_{c2}$). Also, since signatures of the ODT in conventional superconductors get considerably broadened in the presence of random pinning, it has been suggested by some authors that the ODT could be a continuous cross-over rather than a phase transition[13,14,15].

An alternative viewpoint to the thermal route to melting is the disorder induced ODT originally proposed by Vinokur et al.[16]. Here it was speculated that in the presence of weak pinning the transition can be driven by point disorder rather than temperature. In this scenario topological defects proliferate in the VL through the local tilt of vortices caused by point disorder, creating an "entangled solid" of vortex lines. The key difference with conventional thermal melting is that here, the positional entropy generates instability in the ordered VL driving it into a disordered state, even when thermal excitation alone is not sufficient to induce a phase transition. Recently, similar notions have also been extended to the melting of vortex lattice in Bose-Einstein condensates formed of ultra-cold atoms in a disordered optical potential[17].

Real space imaging through scanning tunneling microscopy/spectroscopy (STM/S) is one of the most powerful tool to image the VL over a large range of temperature and magnetic field. Recently, detailed STS imaging of the VL across the ODT in a $Co_{0.0075}NbSe_2$ single crystal[18] revealed that the hexagonal ordered state (OS) of the VL disorders in two steps, reminiscent of the Berezinski-Kosterlitz-Thouless-Halperin-Nelson-Young (BKTHNY) transition[19] in two-dimensional solids. At a fixed temperature as the magnetic field is increased, dislocations, in the form of nearest-neighbor pairs with 5-fold and 7-fold coordination, first proliferate in the VL. We call this state an orientational glass (OG). At a higher field dislocations dissociate into isolated disclinations driving the VL into an amorphous vortex glass (VG). These three states are characterized by their positional and orientational order. In the OS, the VL has long-range or quasi long-range positional order and long-range orientational order. The OG is characterized by a rapid decay of positional order, but a (quasi)-long range orientational order analogous to the hexatic state in 2-dimensional (2D) solids. In the VG both positional and orientational order are short range. (A somewhat different two-step disordering sequence has also been reported in neutron irradiated $NbSe_2$ single crystals[20], though in that case the study was restricted to low fields.) However, since a BKTHNY transition is not expected for a 3-dimensional VL, it is important to investigate whether these two transformations correspond to two distinct phase transitions, or a gradual cross-over as suggested by some authors[13,14].

In this paper, we address the ODT in Co-intercalated $NbSe_2$ by tracking the structural evolution of the VL, imaged using STS. Our data provides structural evidence of superheating and supercooling across both OS-OG and OG-VG transitions. Furthermore, comparing crystals with different degree of pinning, we show that these two transitions come closer to each other when pinning is reduced, suggesting that they are fundamentally associated with the random background potential created by random pinning.

**II. Sample growth, characterization, phase diagram**

The samples used in this study consist of pure and Co-intercalated NbSe$_2$ single crystals[21] grown through iodine vapor transport in sealed quartz ampules. The random intercalation of Co provides us a handle to control the degree of pinning[22]. Single crystals were grown by iodine vapor transport method starting with pure Nb, Se and Co, together with iodine as the transport agent. Stoichiometric amounts of pure Nb, Se and Co (only Nb and Se for the pure sample), together with iodine as the transport agent were mixed and placed in one end of a quartz tube, which was then evacuated and sealed. The sealed quartz tube was heated in a two zone furnace between 5 to 10 days, with the charge-zone and growth-zone temperatures kept at, 800 °C and 720 °C respectively. We obtained single crystals with lateral size (in the a-b plane) of 2-4 mm and typical thickness varying between 60-150 µm. We used three crystals with different levels of pinning for this study. For the first crystal on which all the STS measurements were carried out, we started with a nominal composition Co$_{0.0075}$NbSe$_2$ and the growth was continued for 5 days. We obtained single crystals with narrow distribution of $T_c$, in the range 5.82 – 5.93 K. The crystal chosen for our studies had a $T_c$ ~ 5.88 K (sample S1). The second crystal was also grown starting from the same nominal composition (in a different ampule), but the growth was continued for 10 days. Here we obtained crystals with $T_c$ varying in the range 5.8 – 6.2 K. We conjecture that this larger variation of $T_c$ results from Co gradually depleting from the source such that crystal grown in later periods of the growth run have a lower Co concentration. However, over the 2mm × 2mm crystal chosen from this growth run ($T_c$ ~ 6.18 K) (sample S2), we did not see significant compositional variation. The third sample was a pure NbSe$_2$ single crystal with T$_c$ ~ 7.25 K (sample S3). Compositional analysis of the three crystals was performed using energy dispersive x-ray analysis (EDX). We obtained a Co concentration of 0.45 atomic % for S1 and 0.31 atomic % for S2. While these Co concentrations are marginally higher than the ones reported in ref. 22, we note that these measurements are close to the resolution limit of our EDX machine, where precise determination of the absolute value is difficult. However, measurements at various points on the crystals revealed the composition to be uniform, which is also corroborated by the sharp superconducting transitions observed from a.c. susceptibility in these crystal (Fig. 1(a)).

The bulk pinning properties at low temperature are characterized through d.c. magnetization measurements using a Quantum Design SQUID magnetometer. The magnetic field in all measurements reported in this paper is applied along the *c*-axis of the hexagonal crystal structure. Figure 1 (b) shows the representative hysteresis-loop in the d.c. magnetization (*M-H*) measured at 1.8 K for the sample S1. The hysteresis first collapses below our resolution limit at fields above 5 kOe and then opens up showing a bubble close to $H_{c2}$. This reopening of the hysteresis curve signals a sudden anomalous increase in the critical current, the "peak effect" [23], which is associated with the ODT of the VL. We estimate the critical current ($J_c$) as a function of magnetic field, using the critical state model[24] which relates the width of the hysteresis loop (*ΔM*) to $J_c$ of the superconductor, through the relation, $J_c \approx 20\Delta M/d$ where $J_c$ is in units of A/cm$^2$, *ΔM* is in units of emu/cm$^3$ and *d* is the lateral dimension perpendicular to the applied magnetic field of the crystal in cm. Fig. 1(c) shows *J$_c$(H)* at 1.8 K for the three crystals under investigation. While the absolute value is likely to have some error due to demagnetization effects since the magnetic field is applied along the short dimension of the crystal, we observe that the peak in $J_c$ progressively increases for samples with lower $T_c$ showing that the pinning becomes progressively stronger as Co is introduced. However, we note that even our most disordered sample (S1) is in the weak pinning limit, which is functionally defined as the pinning range where a topologically ordered vortex lattice is realized at low temperatures and low fields.

The phase diagram for the three crystals is established from variation of the real part of a.c. susceptibility with magnetic field ($\chi'$-$H$) at different temperatures in a home built a.c. susceptometer (Fig. 2). We have earlier shown[18] that above 10 mOe of a.c. excitation field the $\chi'$ response becomes non-linear. Here we also observed that in the non-linear regime the onset of the peak effect moves to higher fields with increase in excitation. We believe that this is a consequence of cycling the sample through minor hysteresis loops resulting from the oscillatory a.c. excitation. Consequently in all these measurements we fix the a.c. excitation field to 3.5 mOe (frequency of 31 kHz) which is well within the linear regime. The sample is first cooled in zero magnetic field across the superconducting transition (the zero field cooled (ZFC) protocol) before the magnetic field is applied. $\chi'$-$H$ (*inset* Fig. 2 (a-c)) shows a characteristic drop with magnetic field below $H_{c2}$, the peak effect, signaling the ODT of the VL. Correlating with real space STS images, we earlier identified two characteristic fields[18]: The field at which $\chi'$ starts to decrease (defined as the onset of peak effect, $H_p^{on}$), where dislocations start proliferating into the VL and the peak of the peak effect ($H_p$) where disclinations start proliferating in the VL. Tracking the loci of $H_p^{on}$ and $H_p$ obtained from isothermal $\chi'$-$H$ scans at different temperatures, we obtain two lines in the *H-T* parameter space demarcating the regions of existence of an OS, the OG and the VG. We observe that region in the *H-T* space where we observe the OG gradually shrinks as pinning is reduced. In this context we would like to note that the onset of the peak effect in $\chi'$ is lower than the corresponding onset of the peak effect from d.c. magnetization measurements, a feature which has also been reported in other weakly pinned Type II superconductors[25]. While the reason for this difference is still not completely understood, one likely reason is the inhomogeneity in the superconducting magnet. Unlike χ' measurements, where the sample is held at a fixed position, in conventional d.c. magnetization measurements, the sample in moved over a distance inside the pick-up coils. In this process the sample is cycled through minor hysteresis loops dictated by the inhomogeneity in magnetic field in the superconducting magnet which tends to collapse the magnetic hysteresis when $J_c$ is small[26,27]. $H_p$ on the other hand is the same in both measurements, since the critical current is large at this field and the magnet inhomogeneity has very little effect.

### III. Real space investigations of the VL from STS imaging

We now investigate the thermal history dependence of the VL by imaging the VL using STS imaging across the OS→OG and OG→VG transitions. We focus on the crystal S1 for which these two boundaries are well separated in the *H-T* parameter space. STS measurements are performed using a home built low temperature scanning tunneling microscope[28] operating down to 350 mK and fitted with a 90 kOe superconducting solenoid. Prior to STS measurements the crystal is cleaved in-situ exposing atomically smooth facets in the *a-b* plane, several microns in size. The VL is imaged[29] by measuring the tunneling conductance over the sample surface $\left(G(V) = \left.\frac{dI}{dV}\right|_V\right)$ at fixed bias voltage ($V$ = 1.2 mV) close to the superconducting energy gap, such that each vortex manifest as a local minimum in the tunneling conductance. Topological defects in the VL are identified by first Delaunay triangulating the VL and finding the nearest neighbor coordination for each point. The magnetic field is applied along the six-fold symmetric *c*-axis of the hexagonal NbSe$_2$ crystal.

We first concentrate on the boundary separating the OS and OG. The upper panels of Fig. 3 show the images of the VL acquired at different points during the temperature cycling and the lower panels show the corresponding autocorrelation functions defined as, $G(\bar{r}) = \sum_{r'} f(\bar{r}+\bar{r}') f(\bar{r})$ where $f(\bar{r})$ is the image matrix. A faster radial decay of the auto-correlation function implies a more disordered state. Figure 3 (a) shows the VL in at 8 kOe and 420 mK (corresponding to point **A** in Fig. 2(a)) prepared using ZFC protocol. The VL is in the topological defect free OS. We now heat the sample to 4.14 K (point **B** in Fig. 2(a)) without changing the field, thereby crossing the OS-OG boundary in the phase diagram. However, the VL (Fig. 3 (b)) continues to remain topologically ordered. To demonstrate that this is actually a metastable superheated state, we apply a small magnetic field perturbation in the form of a pulse, by ramping up the field by 300 Oe over 8 sec followed by a dwell time of 5 sec and then ramping down over 8 sec to its original value. It has been shown earlier[13,18] that such a.c. or d.c. magnetic (or current[30,31]) perturbation helps to overcome the local potential barriers causing a dynamic transition from a metastable state towards equilibrium state of the VL. In this case after application of a field pulse, dislocations proliferate in the VL driving it into the OG state (Fig. 3(c)). Subsequent application of field pulse does not alter the state anymore, showing that this is the equilibrium state of the VL. To demonstrate supercooling across the OS-OG transition we heat the crystal to 4.35 K (point **D** in Fig. 2(a)) and cool it back in the same field to 1.6 K (point **C** in Fig. 2(a)). The additional heating is to ensure that the VL completely relaxes in OG state. The supercooled VL continues to have dislocations, characteristic of the OG state (Fig. 3(d)). However, after applying a 300 Oe magnetic field pulse the dislocations annihilate driving the VL into the equilibrium OS (Fig. 3(e)).

We now focus on the OG-VG boundary. For this we prepare the VL in a field of 24 kOe at 420 mK (point **E** in Fig. 2(a)) using ZFC protocol. The VL shown in Fig. 4(a) contains dislocations as expected in the OG state. Fig. 4(b) shows the VL when the crystal is heated above OG-VG boundary to 2.2 K keeping the field unchanged (point **F** in Fig. 2(a)). Here, the number of dislocations greatly increase. In addition to dislocations comprising of nearest neighbor pairs of 5-fold and 7-fold coordinated vortices, we also observe dislocations comprising of nearest neighbor pairs with 4-fold and 8-fold coordination, 8-fold coordinated site with two adjacent 5-fold coordinated site, and 4-fold with two adjacent 7-fold coordinated site. In addition, we also observe a small number of disclinations in the field of view. To determine the nature of this state, we examine the 2-D Fourier transform (FT) of the VL image. The FT of the image shows 6 diffuse spots showing that the orientational order is present in the VL. This is not unexpected since a small number of disclinations do not necessarily destroy the long-range orientational order[32]. This state is thus a superheated OG state. (Further evidence of OG is obtained from the orientational correlation function, $G_6(|r|)$ discussed later.) However, when a magnetic field pulse of 300 Oe is applied large number of disclinations proliferate the VL (Fig. 4(c)) and the FT shows an isotropic ring, corresponding to an amorphous VG. When the crystal is subsequently cooled to 1.5 K (point **G** in Fig. 2(a)) the FT shows an isotropic ring corresponding to a VG. This state is the supercooled VG (Fig. 4(d)). When a magnetic field pulse of 300 Oe is applied at this temperature, the disclinations disappear driving the VL into its equilibrium OG state where the FT recovers the clear six-fold pattern (Fig. 4(e)). Figure 4(f) shows the orientational correlation functions, $G_6(r) = \langle \Psi_6(0) \Psi_6^*(r) \rangle$, which measure the spatial variation of the orientational order parameter, $\Psi_6(r) = \exp[6i\theta(r)]$, where $\theta(r)$ is the angle of a bond between two nearest neighbor points on the lattice located at position $r$, with respect to an arbitrary reference axis[33]. For the superheated OG state at 2.2 K and the equilibrium OG state at 1.5 K, $G_6(r)$ tends towards a constant value for large *r* showing long range orientational order. On the other hand for the supercooled VG state at 1.5 K and the equilibrium VG state at 2.2 K, $G_6(r)$ tends towards zero for large *r*, characteristic of an isotropic amorphous state.

In principle, the OS-OG and OG-VG phase boundaries can also be crossed by isothermal field ramping. However, earlier field ramping measurements[18] performed at 350 mK did not provide unambiguous evidence of superheating and supercooling though a significant hysteresis was observed between the field ramp up and ramp down branch. The most likely reason is that field ramping which changes the density of vortices involves large scale movement of vortices, which provides the activation energy to drive the VL into its equilibrium state. In contrast temperature sweeping does not significantly perturb the vortex lattice owing to the low operating temperatures, and makes these metastable states observable.

In this context it is important to note that for a glassy system the presence of thermal hysteresis alone does not necessarily imply a phase transition, since due to random pinning the VL might not be able to relax to its equilibrium configuration with change in temperature even if we do not cross any phase boundary. We have also observed this kind of metastable states in our experiments. However, the key difference with this kind of metastable states and the superheated/supercooled states is that in this case the difference is merely in the number of topological defects. For example, ramping up the temperature at fixed field within the OG state creates such metastable states which vary from the corresponding equilibrium state only in the number of dislocations (and in the asymptotic value of $G_6(r)$) though both states have long-range orientational order[29]. In contrast, the superheated and supercooled states are distinct from the corresponding equilibrium states both in the nature of topological defects and consequently in their symmetry properties.

**IV. Discussion and summary**

Our experiments provide structural evidence of superheating and supercooling across both OS-OG and OG-VG phase boundaries, a hallmark of thermodynamic phase transition. BKTHNY mechanism of two-step melting is not applicable in this system since it requires logarithmic interaction between vortices which is not realized in a 3-D VL. In our case, the two-step disordering is essentially induced by the presence of quenched random disorder in the crystalline lattice, which provides random pinning sites for the vortices. Further evidence for this is obtained by comparing $\chi'$ as a function of reduced magnetic field, $h = H / H_{c2}$, at 1.7 K for S1, S2 and S3. Fig. 5 (a)-(c) show $\chi'$-$h$ for the three crystals. As the pinning gets weaker the difference $\Delta h = h_p - h_p^{on}$ decreases thereby shrinking the phase space over which the OG state is observed (Fig. 2). We speculate that in the limit of infinitesimal small pinning, $\Delta h \rightarrow 0$, thereby merging the two transitions into a single first order transition possibly very close to $H_{c2}$.

In this context, we can also compare our results with the neutron scattering experiments reported in refs.8 and 13 which studied the thermal history dependence of the VL across the ODT. The results presented there have strong similarities with our results, with the important difference that the two distinct steps accompanying the ODT was not identified. This could be for two reasons. First, the samples used in those studies could have weaker pinning. More importantly, the experiments there were performed in a relatively low field ($H \ll H_{c2}$) where the OG state is observed over a very narrow range of temperatures and is difficult to resolve unless measurements are performed at very small temperature intervals. Evidence of superheating and supercooling of the VL has also been observed from bulk transport measurements[34], though such measurements cannot discriminate between the OG from the VG.

In summary, we provide structural evidence that in the presence of weak pinning, the VL in a 3-D superconductor disorders through two thermodynamic topological phase transitions. This calls for further experimental and theoretical investigations. Experimentally, it would be worthwhile to look for confirmatory signatures of these transitions in thermodynamic measurements such as specific heat, though such signatures are likely to be very weak. Theoretically, it would be interesting to investigate the role of disclinations in the VL, which has not been explored in detail so far. It would also be interesting to explore to what extent these concepts can be extended to other systems such as colloids, charge density waves and magnetic arrays where a random pinning potential is almost always inevitably present.

*Acknowledgement:* The authors would like to thank Valerii Vinokur, Pradeep Kumar, Gabriela Pasquini for critically reading the manuscript and providing valuable input. We also thank Shobo Bhattacharya for stimulating discussions. The authors would like to thank Department of Atomic Energy, Government of India and Science and Engineering Research Board, Government of India for financial support (Grant No. EMR/2015/000083).

[23] G. D'Anna et al., *Flux-line response in 2H-NbSe$_2$ investigated by means of the vibrating superconductor method*, Physica C **218**, 238 (1993).

[24] C. P. Bean, *Magnetization of High-Field Superconductors*, Rev. Mod. Phys **36,** 31 (1064).

[25] S. Kumar et al., *Unveiling of Bragg glass to vortex glass transition by an ac driving force in a single crystal of Yb$_3$Rh$_4$Sn$_{13}$*, Supercond. Sci. Technol. **28**, 085013 (2015).

[26] G. Ravikumar et al., *Effect of field inhomogeneity on the magnetisation measurements in the peak effect region of CeRu$_2$ superconductor,* Physica C **276,** 9 (1997).

[27] For a scan length of 3 cm used in the Quantum Design SQUID, we estimate the magnetic field inhomogeneity experienced by the sample at 15 kOe to be of the order of 9 Oe; see G. Ravikumar et al., *A novel technique to measure magnetisation hysteresis curves in the peak-effect regime of superconductors,* Physica C **298,** 122 (1998).

[28] A. Kamlapure et al., *A 350 mK, 9 T scanning tunneling microscope for the study of superconducting thin films on insulating substrates and single crystals*, Rev. Sci. Instrum. **84**, 123905 (2013).

[29] Section I of the supplementary material gives details of image processing. Section II shows the thermal hysteresis of the VL within the OG state without crossing any phase boundary.

[30] A. E. Koshelev and V. M. Vinokur, *Dynamic Melting of the Vortex Lattice*, Phys. Rev. Lett. **73**, 3580 (1994).

[31] Guohong Li, Eva Y. Andrei, Z. L. Xiao, P. Shuk, and M. Greenblatt, *Onset of Motion and Dynamic Reordering of a Vortex Lattice,* Phys. Rev. Lett. **96**, 017009 (2006).

[32] T. Giamarchi and P. Le Doussal, *Elastic theory of flux lattices in the presence of weak disorder*, Phys. Rev. B **52**, 1242 (1995).

[33] Details of the calculation of $G_6(|r|)$ are given in ref. 18.

[34] Z. L. Xiao, O. Dogru, E. Y. Andrei, P. Shuk, and M. Greenblatt, *Observation of the Vortex Lattice Spinodal in NbSe$_2$*, Phys. Rev. Lett. **92**, 227004 (2004).
15

**Figure Captions:**

**Figure 1.** (a) Temperature variation of $\chi'$ in zero applied d.c magnetic field for samples S1, S2 and S3. (b) 5-quadrant *M-H* loop for the sample S1 at 1.8 K; the inset shows the expanded view of the peak effect. (c) Variation of $J_c$ with magnetic field at 1.8 K for the crystals S1, S2 and S3.

**Figure 2.** (a)-(c) Phase diagrams of pure and Co intercalated NbSe$_2$ crystals (S1, S2 and S3), showing the variation of $H_p^{on}$, $H_p$ and $H_{c2}$ as a function of temperature; the solid lines connecting the points are guides to the eye. Below $H_p^{on}$ the VL is in a topologically ordered state. Between $H_p^{on}$ and $H_p$ the dislocations proliferate in the VL. Between $H_p$ and $H_{c2}$ disclinations proliferate in the VL. The horizontal lines in (a) represent the path along which the hysteresis in the VL is measured. The *insets* show $\chi'(H)/\chi'(0)$ as a function of *H* at 1.7 K; the onset of the peak effect $H_p^{on}$, the peak of the peak effect $H_p$ and the upper critical field $H_{c2}$ are marked with arrows.

**Figure 3.** Hysteresis of the VL across the OS-OG boundary. Conductance maps (upper panel) and the corresponding autocorrelation function (lower panel) showing (a) the ZFC VL created at 0.42 K in a field of 8 kOe; (b) the VL after heating the crystal to 4.14 K keeping the field constant and (c) after applying a magnetic field pulse of 300 Oe at the same temperature; (d) the VL at 1.6 K after the crystal is heated to 4.35 K and cooled to 1.6 K; (e) VL at 1.6 K after applying a magnetic field pulse of 300 Oe. In the upper panels, Delaunay triangulation of the VL is shown with black lines and sites with 5-fold and 7-fold coordination are shown with red and white dots respectively. The color-scale of the autocorrelation functions is shown in the bottom.

**Figure 4.** Hysteresis of the VL across the OG-VG boundary. Conductance map showing (a) the ZFC VL created at 0.42 K in a field of 24 kOe; (b) the VL after heating the crystal to 2.2 K keeping

the field constant and (c) after applying a magnetic field pulse of 300 Oe at the same temperature; (d) the VL after the crystal is subsequently cooled to 1.5 K. (e) VL at 1.5 K after applying a magnetic field pulse of 300 Oe. The right hand panels next to each VL image show the 2D Fourier transform of the image; the color bars are in arbitrary units. Delaunay triangulation of the VL is shown with black lines, sites with 5-fold and 7-fold coordination are shown with red and white dots respectively and sites with 4-fold and 8-fold coordination are shown with purple and yellow dots respectively. The disclinations are circled in green. (f) Variation of $G_6$ as a function of $r/a_0$ (where $a_0$ is the average lattice constant) for the VL shown the panels (b)-(e).

**Figure 5.** (a)-(c) Variation of $\chi'(h)/\chi'(0)$ as a function of reduced magnetic field, $h = H/H_{c2}$, at 1.7 K for three crystals S1, S2 and S3. $T_c$ and $\Delta h$ of the crystals are shown in the legend and $h_p^{on}$ and $h_p$ are marked with arrows.

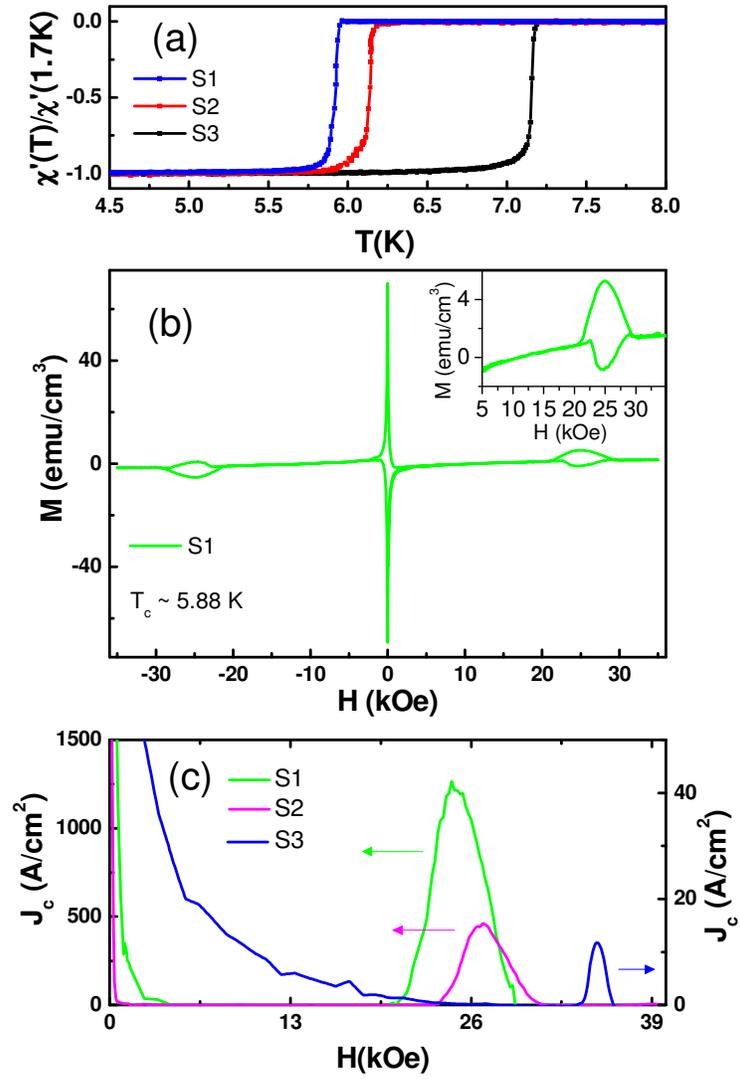

**Figure 1**

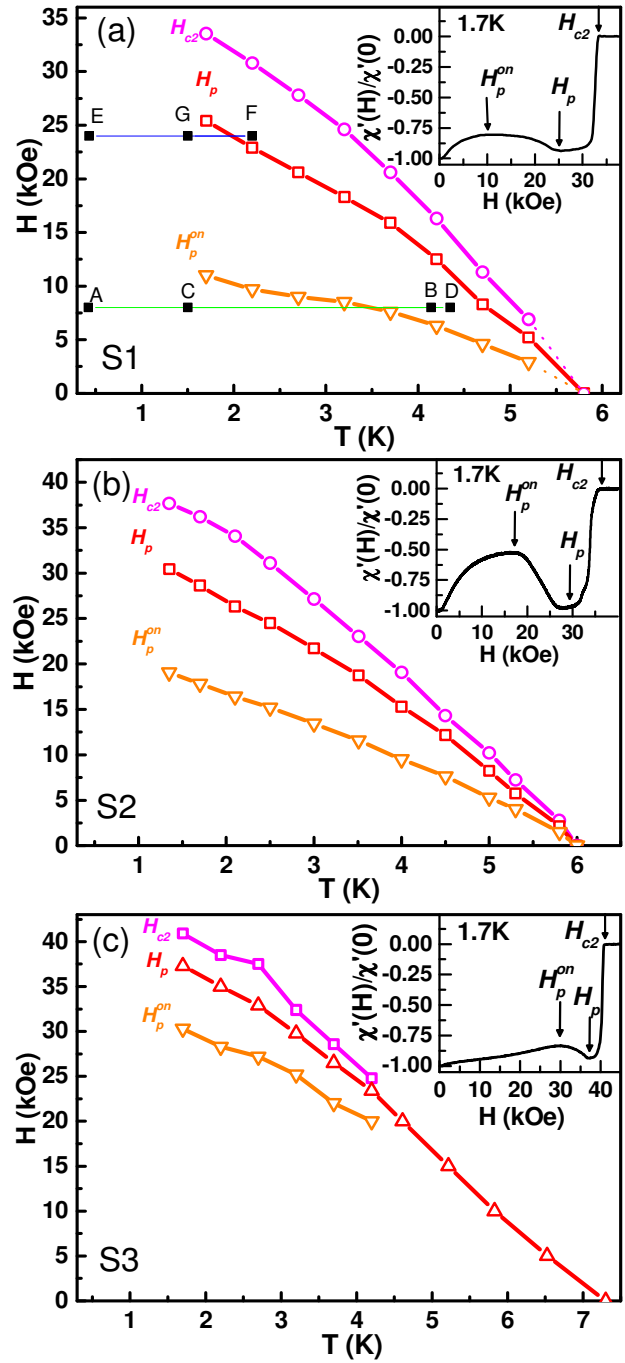

**Figure 2**

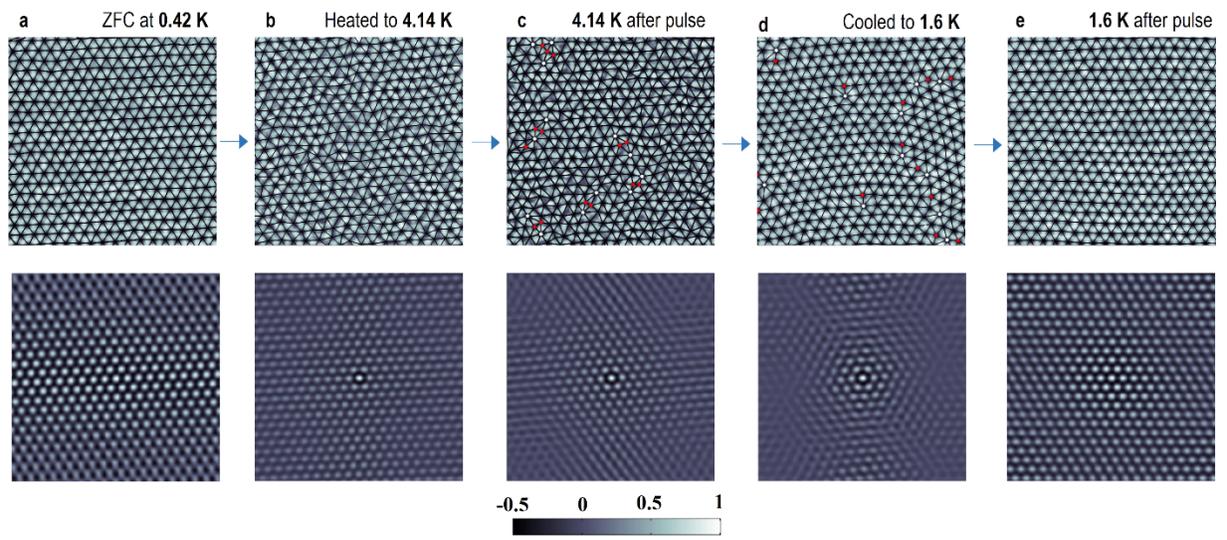

**Figure 3**

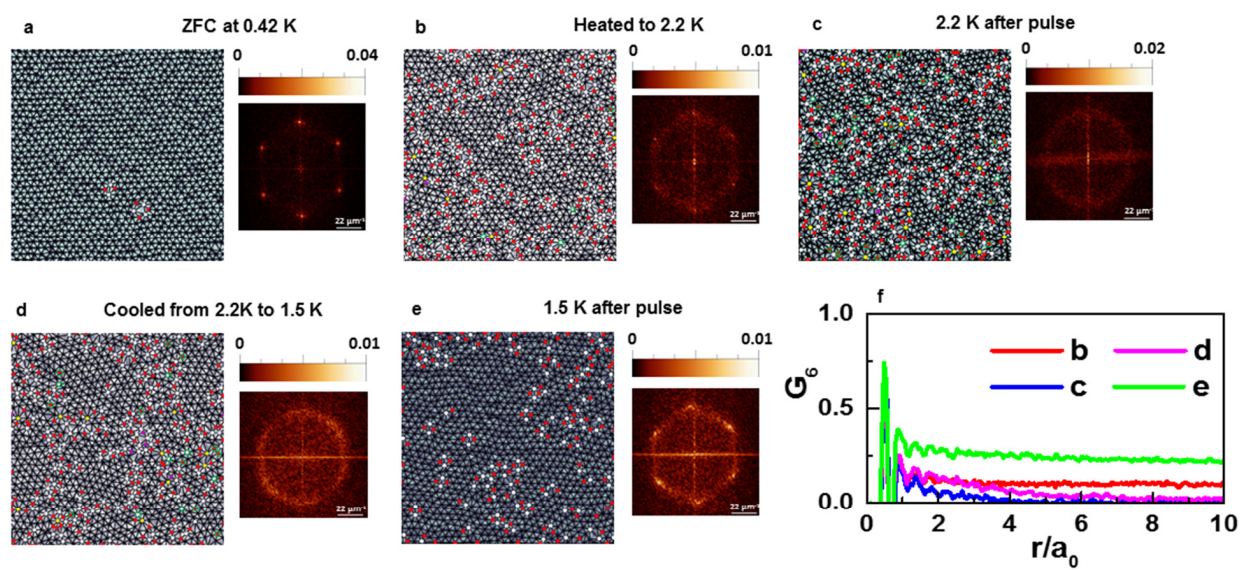

**Figure 4**

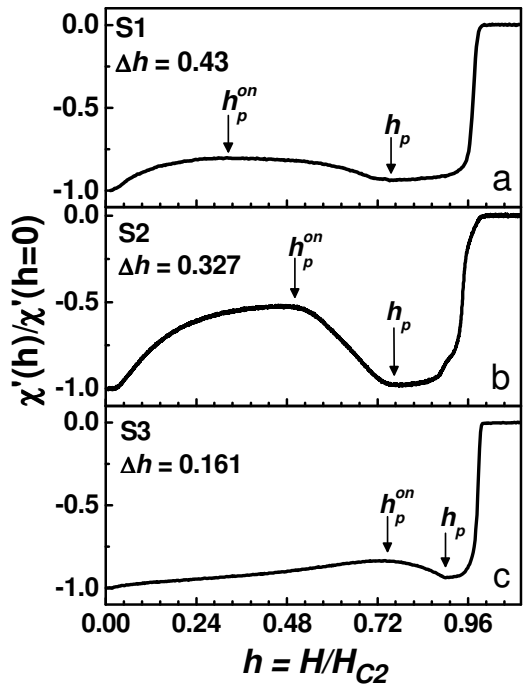

**Figure 5**

**Structural evidence of disorder induced two-step melting of vortex matter in Co-intercalated NbSe₂ single crystals from superheating and supercooling**

Somesh Chandra Ganguli, Harkirat Singh, Indranil Roy, Vivas Bagwe, Dibyendu Bala, Arumugam Thamizhavel and Pratap Raychaudhuri

*Tata Institute of Fundamental Research, Homi Bhabha Road, Colaba, Mumbai 400005, India.*

**I. Filtering the STS conductance maps**

For better resolved vortex image with well-defined minima, the conductance maps obtained from STS are digitally filtered to remove the noise and scan lines which arise from the raster motion of the tip. The filtering procedure is depicted in Fig. S1. Fig. S1(a) shows the raw conductance map obtained at 24 kOe. To filter the image we first obtain the 2D Fourier transform (FT) of the image Fig. S1(b). In addition to six bright spots corresponding to the Bragg peaks we observe a diffuse intensity at large $k$ corresponding to the random noise and a horizontal and a vertical line corresponding to the scan lines. We first remove the noise and scan line contribution from the FT by suppressing the intensity along both the lines and the diffuse intensity at large $k$ (Fig. S1(c)). The filtered image shown in Fig. S1(d) is obtained by taking a reverse FT Fig. S1(c).

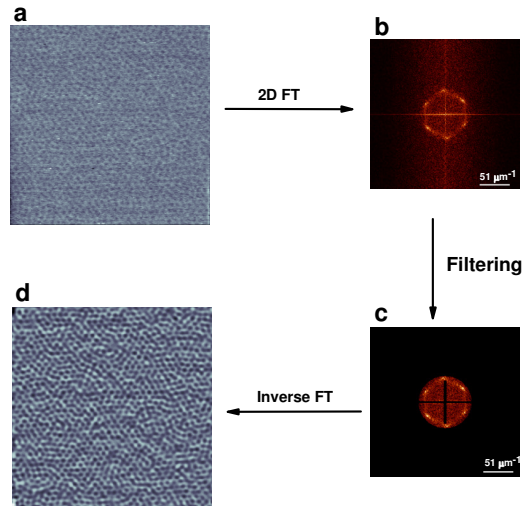

**Fig. S1|** Filtering the conductance map. (a) Raw conductance map recorded at 24 kOe, 1.5 K over an area of 1 μm × 1 μm; (b) 2D FT of (a); (c) 2D FT after removing the contribution from random noise at small $k$ and scan lines; (d) Filtered image obtained from the inverse FT of (c).

## II. Thermal hysteresis of the VL without crossing any phase boundary

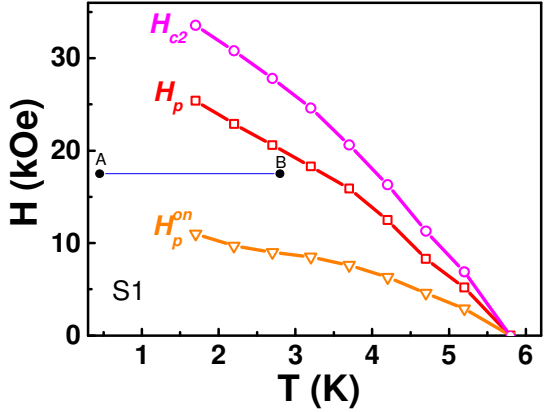

**Figure S2|** Phase diagram of sample S1 showing $H_p^{on}$, $H_p$ and $H_{c2}$.

In the presence of random pinning, thermal hysteresis of the VL can be observed even in the absence of any phase transition. These metastable states result from the inability of the VL to relax to its equilibrium configuration due to the low operating temperatures of conventional superconductors and the presence of random pinning. This kind of metastable states possess the same symmetry as that of the corresponding equilibrium state but vary in the density of topological defects which determine the degree of positional and orientational order.

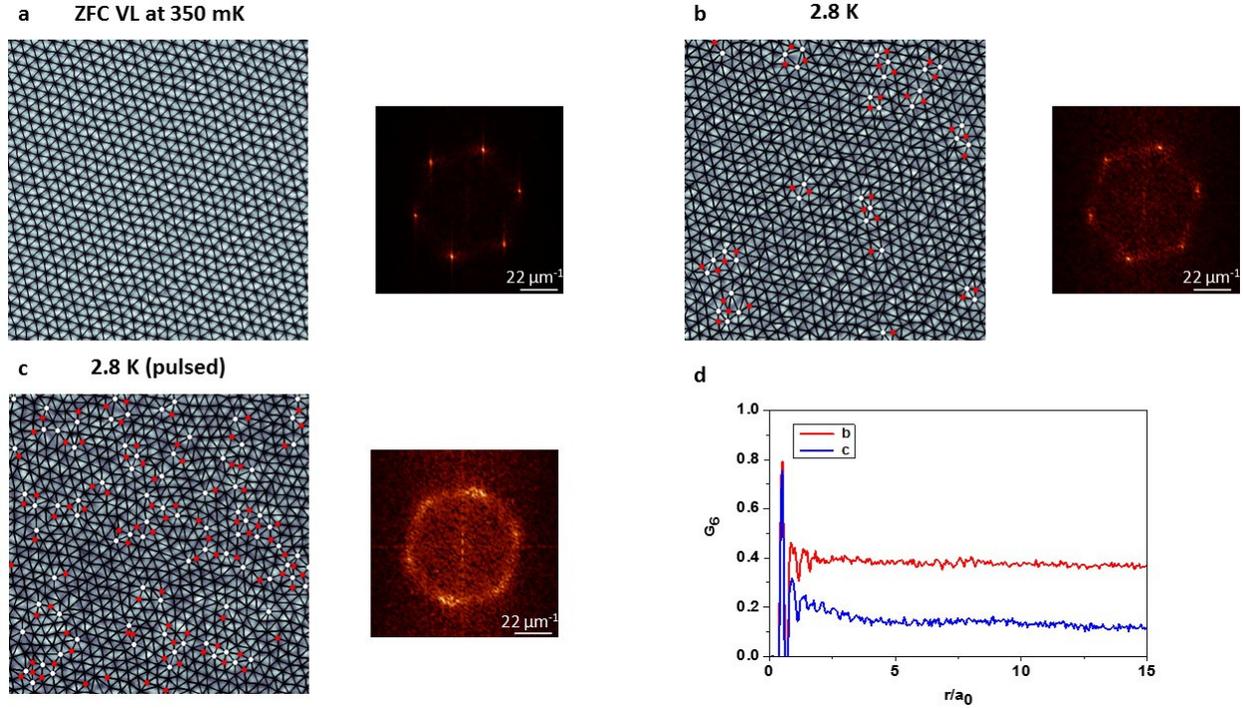

**Fig. S3|** (a) ZFC VL created at 350 mK in a field of 17.5 kOe. (b) The VL after warming up the sample to 2.8 K without changing the magnetic field. (c) VL after applying a magnetic field pulse of 300 Oe. The panels on the right show the Fourier transform of the image. (d) $G_6$ as a function of $r/a_0$ corresponding to the VL in panels (b) and (c) respectively.

In Fig. S3 we show an example of such a metastable state. Fig. S3(a) shows the ZFC VL created in a field of 350 mK in a field of 17.5 kOe (Point **A** in Fig. S2). When the crystal is heated to 2.8 K without changing the field (Point **B** in Fig. S2) we observe a finite density of dislocations (Fig. S3(b)). However, when a magnetic field pulse of 300 Oe is applied on the crystal at the same temperature the density of dislocations greatly increases (Fig. S3(c)), showing that the warmed up ZFC state was an unrelaxed metastable state. However, $G_6(r)$ calculated for the state before and after pulsing shows that both these states have long-range orientational order characteristic of the OG state. Thus the metastable and equilibrium state of the VL here have the same symmetry, while the difference is in the degree of long range orientational order. In contrast the superheated/supercooled states described in the main paper have different symmetry with respect to the corresponding equilibrium states.